\documentclass[10pt,twocolumn,preprintnumbers,amsmath,amssymb,nofootinbib,superscriptaddress]{revtex4-1}

\usepackage{graphicx}
\usepackage{dcolumn}
\usepackage{amssymb,amsmath,bm}
\usepackage{color}
\usepackage[colorlinks,linkcolor=red,citecolor=blue,urlcolor=blue ]{hyperref}
\usepackage{multirow}
\usepackage{comment}
\usepackage{enumitem}
\usepackage{mathptmx} 
\usepackage[utf8]{inputenc}
\usepackage{tensor}
\usepackage{amssymb} 
\usepackage{lettrine}
\usepackage[normalem]{ulem}
\usepackage{empheq}

\newcommand{\bn}{\mathbf{n}}

\newcommand{\dd}{{\rm d}}

\newcommand{\troisj}[6]{\left(\begin{array}{ccc}
      #1 & #2 & #3 \\
      #4 & #5 & #6\end{array}\right)}
\def\bea{\begin{eqnarray}}
\def\eea{\end{eqnarray}}

\def\dd{{\rm d}}

\newcommand{\M}{\mathcal{M}}
\newcommand{\Gal}{_{\rm G}}

\def\AmpGW{{\cal A}}
\def\nn{\nonumber}
\newcommand{\com}[1]{}

\newcommand{\angbr}[1]{\left\langle#1\right\rangle}

\newcommand{\bx}{{\bf{x}}}

\newcommand{\be}{\begin{equation}}
\newcommand{\ee}{\end{equation}}

\newcommand{\gr}[1]{{\bm #1}}

\definecolor{rossos}{cmyk}{0,1,1,0.55}
\definecolor{blu}{cmyk}{1,1,0,0.3}
\definecolor{bluc}{cmyk}{1,1,0,0.1}
\definecolor{verde}{cmyk}{0.92,0,0.59,0.25}
\definecolor{verdec}{cmyk}{0.92,0,0.59,0.15}
\definecolor{verdes}{cmyk}{0.92,0,0.59,0.4}
\hypersetup{colorlinks, bookmarksopen, bookmarksnumbered,
citecolor=blu, linkcolor=bluc, pdfstartview=FitH, urlcolor=rossos}

\definecolor{Mgreen}{rgb}{0.1, 0.69,0.16}

\begin{document}

\title{Measuring anisotropies in the PTA band with cross-correlations}
\author{Giulia Cusin}
\email{cusin@iap.fr}
\affiliation{Institut d'Astrophysique de Paris, UMR-7095 du CNRS et de Sorbonne Universit\'e, Paris, France}
\affiliation{Département de Physique Théorique and Center for Astroparticle Physics, Université de Genève, Quai E. Ansermet 24, CH-1211 Genève 4, Switzerland}
\author{Cyril Pitrou}
\email{pitrou@iap.fr}
\affiliation{Institut d'Astrophysique de Paris, UMR-7095 du CNRS et de Sorbonne Universit\'e, Paris, France}
\author{Martin Pijnenburg}
\email{martin.pijnenburg@unige.ch}
\affiliation{Département de Physique Théorique and Center for Astroparticle Physics, Université de Genève, Quai E. Ansermet 24, CH-1211 Genève 4, Switzerland}
\author{Alberto Sesana}
\email{alberto.sesana@unimib.it}
\affiliation{Dipartimento di Fisica ``G. Occhialini'', Universit\`{a} degli Studi di Milano-Bicocca, Piazza della Scienza 3, I-20126 Milano, Italy}
\affiliation{               INFN, Sezione di Milano-Bicocca, Piazza della Scienza 3, 20126 Milano, Italy}
\affiliation{INAF - Osservatorio Astronomico di Brera, via Brera 20, I-20121 Milano, Italy}

\date{\today}

\begin{abstract}
\noindent
The astrophysical gravitational wave background in the nanohertz (nHz) band is expected to be primarily composed of the superposition of signals from binaries of supermassive black holes. The spatial discreteness of these sources introduces shot noise, which, in certain regimes,  would overwhelm efforts to measure the anisotropy of the gravitational wave background. Moreover, the fact the time-residual map has a finite angular resolution and the presence of pulsar noise, affects our ability to construct the angular power spectrum of the anisotropy from a time-residual map (finite resolution noise). In this work, we explicitly demonstrate, starting from first principles, that cross-correlating a gravitational wave background map with a sufficiently dense galaxy survey can mitigate this issue. This approach could potentially reveal underlying properties of the gravitational wave background that would otherwise remain obscured. We quantify both the shot noise and the finite resolution noise level and show that cross-correlating the gravitational wave background with a galaxy catalog improves by more than one order of magnitude the prospects for a first detection of the background anisotropy by a gravitational wave observatory operating in the nHz frequency range. In particular, we find that with a futuristic scenario with an effective number of frequencies equal to $N_f=10$, the detection of the spectral amplitude can be achieved combining the first $20$ multipoles, with a threshold to resolve single events SKA-like. Increasing observation time, pulsar number or reducing the pulsar white noise considerably improves the detection significance.
\end{abstract}
\maketitle

\section{Introduction}

Pulsar Timing Arrays (PTAs) were used to provide the first evidence of a stochastic gravitational wave (GW) background in the nHz band~\cite{NANOGrav:2023gor, Reardon:2023gzh, EPTA:2023sfo, Xu:2023wog}, with an integrated energy density of $\Omega_{\text{GW}} = 9.3_{-4.0}^{+5.8} \cdot 10^{-9}$ \cite{NANOGrav:2023gor} and evidence of spatial correlation among different pulsar redshifts following the Hellings-Downs (HD) function. Various detection methods are reviewed in \cite{Romano:2016dpx}, and the physical interpretation of the HD correlation is discussed in \cite{Jenet:2014bea, Romano:2023zhb, Grimm:2024hgi, Grimm:2024lfj, Wu:2024xkp}. Methods for mapping the background have also been developed~\cite{Anholm:2008wy, Mingarelli:2013dsa, Depta:2024ykq, Semenzato:2024mtn} (see also \cite{Bernardo:2023jhs, Cusin:2018avf} for a discussion of polarization with Stokes parameters), and it has recently been proposed that a loud source may account for the evidence of anisotropy reported in~\cite{Grunthal:2024sor}. 
The most recent analysis of the HD signature by NANOGrav~\cite{Agazie:2024qnx} is expressed in multipole space (see \cite{Gair:2014rwa,Roebber:2016jzl,Qin:2018yhy,Hotinli:2019tpc,Nay:2023pwu,Bernardo:2022xzl,Bernardo:2022rif,Allen:2024bnk,Bernardo:2024bdc, Pitrou:2024scp} for a harmonic formulation of the HD correlation), with a clear detection of the quadrupole and only marginal evidence for the octupole. The significance of background detection is expected to improve in the future, as the signal-to-noise ratio (SNR) of PTA observables increases with observation time~\cite{Siemens:2013zla,Nay:2023pwu,Pol:2022sjn}. 

The origin of the signal is still uncertain, with anisotropies in energy density providing a key observable to pin down its nature.
In fact, while the overall GW signal generated by an astrophysical population of sources will inevitably show some degree of anisotropy, a cosmological background sourced in the early Universe is expected to be isotropically distributed (at least at angular scales of tens of degrees, probed by PTAs).\footnote{Cosmological backgrounds can in fact exhibit large kinematic anisotropies in the presence of spectral features \cite{Cusin:2022cbb, Cruz:2024svc}, see also \cite{Cruz:2024diu} for detection prospects exloiting sinergies with astrometry.}
Current sky bounds are set by NANOGrav \cite{NANOGrav:2023tcn}, while \cite{Depta:2024ykq} uses Fisher Information Matrix forecasts to show that sensitivity to angular power spectra improves rapidly with the number of pulsars monitored, expected to increase by 20-40 per year.

The astrophysical gravitational wave background (AGWB) in the frequency band targeted by PTA experiments is expected to be primarily dominated by signals from binary systems of massive black holes (BBH) in the inspiraling phase \cite{1995ApJ...446..543R,2003ApJ...583..616J,2008MNRAS.390..192S}. Although the duration of these signals is exceedingly long compared to typical observation times, making the resulting background effectively continuous and stationary over the observation period (but see \cite{2025PhRvD.111b3047F}), the sources themselves have a discrete spatial distribution. The number of sources fluctuates according to a Poisson distribution, introducing shot noise into the angular power spectrum. Additionally, sources are located in galaxies, hence they are expected to be a biased tracer of the large-scale structure distribution. In certain situations, shot noise can obscure or dominate the clustering component, making it challenging to isolate the underlying anisotropic structure of the AGWB spectrum. For example, \cite{Semenzato:2024mtn}  recently   showed using numerical simulations that the auto-correlation map is Poisson-noise dominated, for the black hole models used in the analysis.  

Cross-correlating a shot noise dominated background map with a dense galaxy map has been proved to be an effective method to alleviate shot noise limitation in the context of Earth-based detectors, where the Poissonian nature of sources is both spatial and temporal, due to the very short emission duration  of mergers in band \cite{Jenkins:2019uzp,Jenkins:2019nks, Alonso:2020mva, Cusin:2019jpv, Yang:2023eqi, Alonso:2024knf}. In particular, in \cite{Alonso:2020mva} it is  shown, in the context of ground-based detectors and considering an ideal scenario where shot noise is the only noise component, that the SNR of the cross-correlation outperforms that of the auto-correlation by several orders of magnitude.

In the context of PTAs, in addition to shot noise, there exists an additional noise component stemming from the finite angular resolution of the time-residual map. This resolution is determined by the number of pulsars in the network, characterized by \( L_{\text{max}} \sim \sqrt{N_p} \). Even in a perfect experiment with zero pulsar noise, this limitation impacts our ability to constrain the anisotropy of the time-residual map, as discussed in \cite{Hotinli:2019tpc}. Furthermore, it propagates into the reconstruction of the angular power spectrum of the background energy density, which is the primary observable used to quantify deviations from isotropy, see \cite{Pitrou:2024scp}. We will refer to this additional noise component as finite resolution noise. The size of the finite resolution noise is set by the number of effective frequencies which for a perfect experiment (where pulsars are noiseless) is extremely large ($N_f=2 T f_{\text{max}} \gg 1$) while it is much suppressed in the presence of pulsar noise \cite{Pitrou:2024scp}.  Our aim here is to explore the issues of shot noise and finite resolution noise in the context of PTA observations, and to show that cross-correlating a background map with a dense galaxy catalogue can mitigate both noise contributions.

After having identified the observables we are interested in in section \ref{observables}, we present a first principle derivation of the shot noise contribution of auto- and cross-correlation maps for an AGWB in the nHz band. We need to account for two distinct layers of stochasticity describing the distribution of massive black hole binaries: a given galaxy may or not contain a massive black hole binary, and galaxies are themselves a discrete and random biased sampling of the underlying matter density. Using properties of compound statistics, we show that while the shot noise of the auto-correlation map is proportional to the inverse number of sources (black hole binaries), the cross-correlation shot noise is proportional to the inverse galaxy number, hence it is much suppressed. In section \ref{Map noise} we turn into the computation of the finite resolution noise, for both auto- and cross-correlation angular spectra, adapting results from \cite{Hotinli:2019tpc, Pitrou:2024scp}. In section \ref{application} we present numerical results: we consider the theoretical framework of \cite{Cusin:2017mjm, Cusin:2017fwz, Pitrou:2019rjz} to describe anisotropies and we consider the astrophysical models of \cite{2013MNRAS.433L...1S,Rosado:2015epa} to describe the underlying massive black hole population. The AGWB strain
map is dominated by a few very bright sources which obscure a background from an unresolved population,  whose anisotropy map is a representative sample of large scale structure anisotropies. We therefore consider the effect of filtering out bright sources, for three different threshold values of source strain:  a threshold corresponding to the EPTA strain curve ("no cut" situation as effectively there are no resolvable sources resulting from the filtering procedure), a threshold  corresponding to the projected SKA strain sensitivity, and an intermediate value. For each situation we compute the auto and cross-correlation spectrum with galaxy distribution, and the corresponding shot noise and finite resolution noise. We estimate the SNR of auto- and cross-correlation maps and we show that the detectability depends on two factors: on the threshold for filtering out resolvable events, which depends on the sensitivity of the instrument, and on the number of effective frequencies one observatory has access to $N_f$, which depends on the number of pulsars monitored, on the time of observation and on the pulsar noise level, see section VII C 2 of \cite{Pitrou:2024scp}. In that reference, it was actually found that the number of effective frequency bins for fixed observational setting, is a decreasing function of the multipole $L$ of the time residual map. Here we consider an effective  flat number  $N_f\sim 10$, which has to be understood as the average $N_f$ over the number of multipoles one has access to with a given map resolution. The number $N_f=10$ corresponds to a futuristic scenario\footnote{Assuming the pulsar noise characteristics of~\cite{NANOGrav:2023hde}, getting $N_f(L=5)\sim 1$  would require 450 pulsars monitored during
25 years. }, which can only be achieved by considerably increasing observation time and the cadence of pulsar monitoring. We also consider the case of a perfect experiment with $N_f=2 T f_{\text{max}}\gg 1$. In this latter case, the finite resolution noise is very small (pulsar white noise is negligible) and shot noise dominates. For $N_f=10$, we find that detection can be achieved with cross-correlation, integrating the background signal over frequency and combining the first multipoles up to $\ell=20$ with a threshold for resolvable events at $10^{-16}$. For $N_f\gg 1$ detectability is obtained combining the first five multipoles for threshold $10^{-16}$ and the first 13 multipoles for $10^{-15}$.  In this article, technical derivations are presented in a series of appendices.

{\bf{Notation}}. The index $L$ denotes  the angular decomposition of a time-residual map. The angular resolution of the map is related to the number of pulsars in the network via $L_{\text{max}}\sim \sqrt{N_p}$. The lower case $\ell$ is used for the multipole decomposition of the energy density correlation function, and $\ell_{\text{max}}= 2 L_{\text{max}}$. We call \emph{perfect experiment} an observational setting where the pulsar noise is negligible. Hence a perfect experiment still is characterised by a finite pulsar number, observation time, and cadence of observation. In our analysis, the entire redshift distribution of galaxies is utilized, employing optimal Wiener weights, hence assuming that the redshift distribution of GW sources is known.

\section{General concepts and derivations}
\subsection{Our observables}\label{observables}

The observed GW energy density parameter, $\Omega_{\text{GW}}$ is defined as the background energy density $\rho_{\text{GW}}$ per units of logarithmic frequency $f$ and solid angle $\bf{n}$, normalized by the critical density of the Universe today $\rho_c$. It can be divided into an isotropic background contribution $\bar{\Omega}_{\text{GW}}$ and a contribution from anisotropic perturbations $\delta \Omega_{\text{GW}}$ \cite{Cusin:2017fwz, Cusin:2017mjm, Cusin:2019jpv}:
	\begin{align}\label{Splitbackgroundperturbations}
	    \Omega_{\text{GW}}(\bn,f)=\frac{f}{\rho_c}\frac{\dd^3 \rho_{\text{GW}}}{\dd^2 \bn\, \dd f}(\bn,f)=\frac{\bar{\Omega}_{\text{GW}}(f)}{4\pi}+\delta \Omega_{\text{GW}}(\bn,f),
	\end{align}
	where the isotropic background spectrum can be written as the integral over conformal distance $r$ (where we set the speed of light equal to unity,  $c=1$):
    \begin{subequations}
	\begin{align}
	    &\bar{\Omega}_{\text{GW}}(f) = \int \dd r \, \partial_r \bar{\Omega}_{\text{GW}}(f,r),\label{Omegaf0}
	\end{align}
    \end{subequations}
where the integrand plays the role of  an astrophysical kernel that contains information on the local production of GWs at galaxy scales. This Kernel can be built out of a model of black holes formation, or a catalog implementing such model, as detailed in Appendix~\ref{AppC}. Different astrophysical models give quite different predictions for this kernel, see e.g.~\cite{Cusin:2019jpv} for an explorative approach in the frequency band of earth-based and space-based detectors.

    We now consider observables integrated over the whole frequency range a given observatory is sensitive to. Explicitly, we introduce
    \begin{subequations}
    \begin{align}
    \bar{\Omega}_{\text{GW}}&=\int \dd \log f \, \bar{\Omega}_{\text{GW}}(f)\,,\label{OmegaInt0}\\
   \Omega_{\text{GW}}(\bn)&=\int \dd \log f \, \Omega_{\text{GW}}(\bn, f)\,,\\
 \partial_r\bar{\Omega}_{\text{GW}}(r)  &= \int \dd \log f \,\partial_r\bar{\Omega}_{\text{GW}}(f, r)\,,\label{OmegaInt}\\
       \delta\Omega_{\text{GW}}(\bn)&=\int \dd \log f  \,\delta\Omega_{\text{GW}}(\bn, f)\,. \label{dOmegaInt}
    \end{align}
    \end{subequations}
We observe that since $\delta\Omega_{\text{GW}}$ is a stochastic quantity, it can correlate with other cosmological stochastic observables. An interesting observable to look at is the cross-correlation of the AGWB with the distribution of galaxies, i.e. with the galaxy number counts $\Delta$ defined as the overdensity of the number of galaxies per unit of redshift and solid angle
\be\label{Ncounts}
\Delta(\bn, z)\equiv \frac{N(z, \bn)-\bar{N}(z)}{\bar{N}(z)}\,.
\ee
First, if astrophysical GW sources are located in galaxies, we would expect the SGWB and the galaxy distribution to have a high correlation level. Second, cross-correlating with galaxies helps to mitigate the problem of shot noise and to possibly extract the clustering information out of the shot noise threshold. This has been shown in the context of ground-based detectors, where shot noise is dominated by its popcorn component due to the transient emission of sources, see \cite{Alonso:2020mva, Cusin:2019jpv, Jenkins:2019uzp, Jenkins:2019nks}.

We want to show that this is also the case when the emission is continuous in time, like in the PTA band. Finally, by cross-correlating with the galaxy distribution at different redshifts, one could try to get a tomographic reconstruction of the redshift distribution of sources. In this work, our goal is to maximize the detection chances hence we do not follow this approach. 

The angular power spectrum of the GW and galaxy counts auto-correlations and for their cross-correlations are defined as
\begin{subequations}
\begin{align}
(2\ell+1)\,C^{\text{GW}}_{\ell}&\equiv \sum_{m=-\ell}^{\ell} \langle a_{\ell m}\,a^*_{\ell m}\rangle\,,\label{ClGW}\\
(2\ell+1)\,C^{\Delta}_{\ell}&\equiv \sum_{m=-\ell}^{\ell} \ \langle b_{\ell m} \,b^*_{\ell m}\rangle\,,\\
(2\ell+1)\,C^{\text{GW}\Delta}_{\ell}&\equiv \sum_{m=-\ell}^{\ell} \ \langle a_{\ell m}\,b^*_{\ell m}\rangle\,,
\end{align}
\end{subequations}
where the bracket denotes an ensemble average and $a_{\ell m}$ and $b_{\ell m}$ are the coefficients of the spherical harmonics decomposition of the AGWB energy density and galaxy number counts, respectively. Explicitly 
\begin{subequations}
\begin{eqnarray}
\label{Eq:Sec2Sph}
   \delta \Omega_{\text{GW}}(\bn) & = & \sum_{\ell=0}^{\infty} \sum_{m=-\ell}^{\ell} a_{\ell m} \, Y_{\ell m}(\bn)\,,\\
    \Delta(\bn) & = & \sum_{\ell=0}^{\infty} \sum_{m=-\ell}^{\ell} b_{\ell m} \, Y_{\ell m}(\bn)\,.\label{blm}
\end{eqnarray}
\end{subequations}

It can be shown that the angular power spectra of the auto- and cross-correlation are given by \cite{Cusin:2017fwz}: 
\begin{subequations}
	\begin{align}\label{eq:modelCl}
 	    C_{\ell}^{\text{GW}}&= \frac{2}{\pi}\int \dd k\ k^2 \, |\delta \Omega_{\text{GW}\,,{\ell}}(k)|^2\,,\\
      	    C_{\ell}^{\Delta}&= \frac{2}{\pi}\int \dd k\ k^2 \, |\Delta_{\ell}(k)|^2\,,\\
	    C_{\ell}^{\text{GW}\Delta}&= \frac{2}{\pi}\int \dd k\ k^2 \, \delta \Omega^{*}_{\text{GW}\,,{\ell}}(k) \, \Delta_{\ell}(k)\,,
	\end{align}
    \end{subequations}
	where $k$ is the Fourier mode norm. Keeping only the leading-order contribution to the anisotropy given by clustering, which is expressed by~\eqref{eq:deltaOmega}, we have 
		\begin{align}\label{eq:deltaOmegaGW}
	      \delta \Omega_{\text{GW}\,,\ell}(k)&=\int \dd\log f\,\delta \Omega_{\text{GW}\,,\ell}(k,f)\,,\\
          \delta \Omega_{\text{GW}\,,\ell}(k,f)&= \int \frac{\dd r}{4\pi} \, \partial_r \bar{\Omega}_{\text{GW}}(f,r) \, \big[b(r)\,\delta_{m,k}(r)j_{\ell}(k r) \big]\,,
	\end{align}
where $j_{\ell}$ are spherical Bessel functions, while $\delta_{m}$ is the dark-matter over-density, related to galaxy overdensity via the bias factor $b(r)$.  
The corresponding contribution from galaxy overdensities (retaining only the dominant clustering contribution) reads 
		\begin{align}\label{eq:DeltaGC}
	    \Delta_{\ell}(k)=\int \dd r \, W(r)\, \big[b(r)\,\delta_{m,k}(r)j_{\ell}(k r) \big]\,, 
	\end{align}
	where $W(r)$ is a window function normalized to one which selects the redshift bin in the galaxy catalog we want to consider in the cross-correlations.

The framework presented so far is general and can be applied to any astrophysical background component across any frequency band. We will now focus on the specific case of PTAs, where the sources are binary systems of massive black holes residing in galaxies. In the next section, we will estimate the contribution of shot noise in this context, both for the auto-correlation spectrum and for the cross-correlation with the galaxy distribution.

Recently, the variance on the background energy density due to the discrete distribution of sources has been studied in \cite{Lamb:2024gbh}, while \cite{Sato-Polito:2023spo} looks at  
anisotropies due to discrete source distribution, for different astrophysical models.

\subsection{Shot noise}

 Our goal is to demonstrate from first principles that if the source distribution is a sub-sample of the underlying galaxy distribution, cross-correlations are less affected by shot noise than auto-correlations. We will also derive explicit expressions for both contributions.

\subsubsection{Compound statistics}

When counting the number of massive BBH, we have two levels of stochasticity: one galaxy may or may not contain a black hole binary, and the distribution of galaxies is stochastic and modeled as Poisson-distributed. We write the number of black hole binaries as
\begin{align}
N_{\text{BBH}}&=\sum_{i=1}^{N_{\text{G}}}X_i\,,
\end{align}
where $X_i$ follows a binomial distribution $X_i\sim B(1, \beta_X)$, hence $\langle X_i\rangle=\beta_X$ and $\text{Var}(X_i)=\beta_X(1-\beta_X)$. In practice $\beta_X\ll 1$, so that most galaxies do not contain a binary emitting in the relevant frequency band.

 Using properties of compound statistics derived in Appendix \ref{Appcompound} and recalling that the covariance of two variables $Y$ and $Z$ is defined as $\text{Var}(Y, Z)=\langle Y Z\rangle -\langle Y\rangle \langle Z\rangle$, one finds 
\begin{subequations}
\begin{align}
\langle N_{\text{BBH}}\rangle&=\beta_X \langle N_{\text{G}}\rangle\,,\\
\text{Var}(N_{\text{BBH}})&=\beta_X \langle N_{\text{G}}\rangle\,,\\
\text{Var}(N_{\text{BBH}}, N_{\text{G}})&= \beta_X \langle N_{\text{G}}\rangle\,.\label{nont}
\end{align}
\end{subequations}
We observe that $N_{\text{BBH}}$ is also a Poissonian variable.

\subsubsection{A heuristic derivation}\label{heuristic}

Let us try to have an intuition of why cross-correlating the stochastic background map to a galaxy map helps with the shot noise problem. To present a heuristic derivation, we simplify our treatment and notation: we fix a frequency bin, we consider a set of radial bins and we assume that all sources have the same luminosity $L$. 
Recalling that only objects in the same distance bin correlate, we can schematically write 
\be
\Omega_{\text{GW}}=\sum_r \Omega_{\text{GW}}(r) = \sum_r N_{\text{BBH}}(r)\frac{L}{\rho_c 4 \pi r^2}\,, \label{OmegaSum}
\ee
where $L/(4\pi r^2)$ is the energy flux associated to the GW event. For the galaxy distribution, we write
\begin{subequations}
\begin{align}
N_{\text{G}}&=\sum_r N_{\text{G}}(r)\,,\\
\Delta&=\sum_r \frac{N_{\text{G}}(r)-\langle N_{\text{G}}\rangle}{\langle N_{\text{G}}\rangle}\,.
\end{align}
\end{subequations}
Then using results from previous section, one has 
\begin{subequations}
\begin{align}
\text{Var}(\Delta)&=\frac{1}{\langle N_{\text{G}}\rangle^2}\sum_r \text{Var}\left(N_{\text{G}}(r)\right)=\frac{1}{\langle N_{\text{G}}\rangle}\,,\\
\text{Var}(\Omega_{\text{GW}})&=\sum_r \left(\frac{L}{\rho_c 4 \pi r^2}\right)^2 \text{Var}\left(N_{\text{BBH}}(r)\right)\nn\\
&=\sum_r \frac{\langle\Omega_{\text{GW}}(r)\rangle^2}{\langle N_{\text{BBH}}(r)\rangle}\,,\\
\text{Var}(\Omega_{\text{GW}}, \Delta)&=\frac{1}{\langle N_{\text{G}}\rangle}\sum_r \frac{L}{\rho_c 4 \pi r^2}\text{Var}\left(N_{\text{BBH}}(r), N_{\text{G}}(r)\right)\nn\\
&=\frac{\langle \Omega_{\text{GW}}\rangle}{\langle N_{\text{G}}\rangle}\,.
\end{align}
\end{subequations}
Note that $\text{Var}(\Omega_{\text{GW}})$ is suppressed by the number of massive black hole binaries (at fixed $\Omega_\text{GW}$). Since $\langle N_{\text{G}}\rangle \gg \langle N_{\text{BBH}}\rangle$ (for $\beta_X \ll 1$), the relative cross-correlation variance is expected to be smaller than the auto-correlation.

\subsubsection{A more formal derivation of shot noise}

Let us refine the derivation above, giving up the assumption that sources have all the same luminosity. We consider a binned catalog of sources (in our case massive black holes in binaries) and look at their  correlation function. Define $\Delta_{i, f, \M}$ to be the source overdensity in a spatial bin $i$, frequency bin $f$, and chirp mass bin $\M$. The two point function of this observable $\langle \Delta_{i, f, \M} \Delta_{j, f', \M'}\rangle$ is given by two terms 
\be
\langle \Delta_{i, f, \M} \Delta_{j, f', \M'}\rangle=\frac{1}{V_c \Delta \log f \Delta \log \M  a^3\bar{n}_s}\delta_{f f'}\delta_{\M \M'}\delta_{ij}+\xi^{\rm s}_{ij}\,,
\ee
where $\xi^{\rm s}_{ij}$ is the binned correlation function of the underlying, smooth, density field, $\delta_{\rm s}$, and $V_c \Delta \log f \Delta \log \M  a^3 \bar{n}_s$ is the mean number of sources in a spatial pixel of comoving volume $V_c$, frequency pixel of volume $\Delta \log f$, and chirp mass pixel $\Delta \log \M$.  Finally, $\delta_{ij}$ is the Kronecker symbol which arises from the Poisson sampling of the underlying, smooth density field, while the Kronecker in frequency and chirp mass space comes from the fact that only sources emitting at the same frequency are correlated. If we take the continuous limit of a binned survey, we find 
\begin{align}
&\Delta_{i, f, \M}\rightarrow \delta_{\rm s}(\bx, \log f, \log\M)\,,\nn\\
&\delta_{ij}\rightarrow V_c\delta^{(3)}(\bx-\bx')\,,\nn\\
&\delta_{f f'}\rightarrow \Delta \log f \,\delta(\log f-\log f')\,,\nn\\
&\delta_{\M \M'}\rightarrow \Delta \log \M \,\delta(\log \M-\log \M')\,. 
\end{align}
It is convenient to split the observed source overdensity into a smooth clustering part $\delta_{\rm s}(\bx, \log f, \log \M)$ and Poissonian contribution 
\be
\hat{\delta}_s=\delta_s(\bx, \log f, \log \M)+\delta_s^{\text{shot}}(\bx, \log f, \log \M)\,,
\ee
where with a hat we denote observable quantities and the arguments on the left hand side are understood for shortness. The first contribution on the right hand side is due to clustering while for the Poissonian contribution one has 
\begin{align}
&\angbr{\delta_s^{\text{shot}}(\bx, \log f, \log \M)\delta_s^{\text{shot}}(\bx', \log f ', \log \M')}\nn\\
&=\frac{1}{a^3 \bar n_s}\delta^{(3)}(\mathbf x -\mathbf x') \delta(\log f -\log f ')\delta(\log\M -\log\M ')\,,\label{Eq:shot_2point_masses}
\end{align}
where the quantity at the denominator $n_s=n_s(z,\log f, \log\M)$ is the density of binaries per units $\log f$ and $\log \M$. 
We use this result to compute the correlation function of the background anisotropies accounting for both clustering and Poissonian components.
We assume in first approximation that the clustering of sources follows the clustering of galaxies i.e. $\delta_s(\bx, \log f, \log \M)=\delta_G(\bx)$.\footnote{Notice that this is just an approximation and a more refined model should e.g. consider that in fact massive galaxies are more likely to host black holes in binaries \cite{2014MNRAS.439.3986R}. However this would not change the result for shot noise, whose computation is the main goal of our treatment.}
We get 
\begin{widetext}
\begin{align}
&\langle \delta\hat{\Omega}_{\rm GW}(\bn) \delta\hat{\Omega}_{\rm GW}(\bn')\rangle=\frac{1}{(4\pi)^2}\int \dd r \int \dd \log f \int \dd \log \M  \frac{\partial \bar \Omega_{\rm GW}}{\partial r \partial \log \M}\int \dd r' \int \dd \log f' \int \dd \log \M'\frac{\partial \bar \Omega_{\rm GW}}{\partial r' \partial \log \M'}\nn\\
&\times \left[\langle \delta_{\Gal}(\bx=\bn r)\delta_{\Gal}(\bx'=\bn' r')\rangle  +\delta^{(3)}(\bx-\bx')\delta(\log f-\log f') \delta(\log \M-\log \M')\frac{1}{a^3 \bar{n}_{\rm s}}\right]\,,
\end{align}
\end{widetext}
where we made use of~\eqref{dOmegaInt} and \eqref{eq:deltaOmega}. 
Expanding in terms of spherical harmonics
\be
 \delta\hat{\Omega}_{\rm GW}(\bn)=\sum_{\ell m} \hat{a}_{\ell m} Y_{\ell m}(\bn)\,,
 \ee
and using 
\be
\delta^{(3)}(\bx-\bx') = \frac{\delta(r-r')}{r^2} \sum_{\ell m} Y_{\ell m}(\bn) Y_{\ell m}^\star(\bn')\,,
\ee 
we find that 
\begin{align}
&\langle  \hat{a}_{\ell m} \hat{a}^*_{\ell' m'}\rangle=\langle  a_{\ell m}  a^*_{\ell' m'}\rangle+\delta_{mm'}\delta_{\ell \ell'}\times\\
& \int \frac{\dd r}{(4\pi)^2} \int \dd \log f \int \dd\log \M\, \Big|\frac{\partial \bar \Omega_{\rm GW}}{\partial r \partial \log \M}\Big|^2 \frac{1}{r^2} \frac{1}{a^3 \bar{n}_{\rm s}}\,,\nn
\end{align}
where the first contribution is the theoretical one, given in eq.\,(\ref{ClGW}) and the second the shot noise component. Using the standard definition 
\be
\langle  \hat{a}_{\ell m}  \hat{a}^*_{\ell' m'} \rangle\equiv{\hat C}_{\ell}\delta_{\ell\ell'}\delta_{mm'}\,,
\ee
we immediately find that
\be\label{qq}
\hat{C}_{\ell}=C^{\rm GW}_{\ell}+N_{\text{shot}}^{\rm GW}\,,
\ee
where 
\begin{align}\label{shots}
&N_{\text{shot}}^{\rm GW}\equiv\\\
&\int \frac{\dd r}{(4\pi)^2} \int \dd \log f \int \dd\log \M \,\Big|\frac{\partial \bar \Omega_{\rm GW}}{\partial r \partial \log\M}\Big|^2 \frac{1}{r^2} \frac{1}{a^3 \bar{n}_{\rm s}}\nn\,.
\end{align}
One can repeat the same reasoning for the cross-correlation and the galaxy auto-correlation. Using results of the heuristic argument, we can directly see that 
\begin{subequations}
\begin{align}
        N_{\rm shot}^{\text{GW}\Delta}&=\int\frac{\dd r}{4 \pi r^2}\frac{1}{a^3 \bar{n}_G(r)} W(r) \, \partial_r \bar{\Omega}_{\rm GW}(r)\,,\\
        N_{\rm shot}^{\Delta}&=\int\frac{\dd r}{r^2}\frac{1}{a^3 \bar{n}_{G}(r)} W^2(r)\,,
    \end{align}
    \end{subequations}
which scale as $1/(a^3\bar{n}_{\text{G}})$ which is the inverse of the comoving number density of galaxies. In the Appendix\,\ref{Appcatalog}, we explain how the shot noise contribution for the GW auto-correlation can be estimated from a catalog, and we prove the equivalence between the discrete and continuous descriptions. 

To conclude: we found that while shot noise of the AGWB energy density auto-correlation scales as the inverse number of sources (black hole binaries), the shot noise of the cross-correlation with the galaxy distribution scales as the inverse number of galaxies, hence shot noise of a cross-correlation map is much more suppressed than shot noise of the AGWB auto-correlation.  

\subsection{Noise due to finite resolution of the time-residual  map}\label{Map noise}

Even for a perfect experiment (i.e. for zero pulsar noise), there is an additional noise component that affects our ability to reconstruct the angular power spectrum of the background intensity (and its cross-correlation with the large scale structure), due to the limited angular resolution of a given PTA observatory. Indeed the angular power spectrum is built from a map of anisotropies, and the minimum anisotropy that can be detected is set by the map resolution $L_{\text{max}}$, in turn related to the pulsar number $L_{\text{max}} \sim \sqrt{N_p}$, see \cite{Hotinli:2019tpc} for a detailed derivation.    
This in turn propagates on our ability to construct the angular power spectrum. A detailed 
derivation of this can be found in section VE of \cite{Pitrou:2024scp}, and summarised in appendix \ref{NewApp}. The resulting noise is given by 
\begin{align}\label{DefNlGWLmax}
&N^{\text{GW}}_\ell(L_{\text{max}})\nonumber\\
&=\sqrt{\frac{2}{2 \ell+1}}\left(\frac{\bar{\Omega}_{\text{GW}}}{4\pi}\right) \mathcal{B}_{\ell m}
\left[\left(\frac{\bar{\Omega}_{\text{GW}}}{4\pi}\right)^2  \mathcal{B}^2_{\ell m}+2  C_\ell^{\text{GW}}\right]^{1/2}\,,
 \end{align}
where
\begin{align}\label{varA}
&\mathcal{B}^{-2}_{\ell m}\\
&=\frac{N_f}{8\pi}\sum_{L_1 L_2 }\delta_{L_1+L_2+\ell}^{\text{even}}(2L_1+1)(2L_2+1) \troisj{\ell}{L_1}{L_2}{0}{-2}{2}^2\nonumber\,,
\end{align}
where $\delta_{L_1+L_2+\ell}^{\text{even}}\equiv\left(1+(-1)^{L_1+L_2+\ell}\right)/2$ and $N_f$ is equal to the number of effective frequencies, which for a perfect experiment is given by $N_f = 2 T f_{\rm max}$. In a realistic experiment, $N_f$ is suppressed and depends on the pulsar noise level (beside pulsar number and
observation time), see \cite{Pitrou:2024scp} for a detailed discussion.  In all these expressions $0\leq \ell \leq 2 L_{\text{max}}$ and $L_1, L_2 \leq L_{\text{max}}$ with $L_{\text{max}}\sim \sqrt{N_p}$, and $N_p$ is the number of pulsars in the network.

The finite resolution of the GW map also affects the reconstruction of the angular power spectrum of cross-correlation. Assuming the galaxy map to be noiseless (being its angular resolution much better than the GW background one), one finds (see appendix \ref{NewApp}) 
\begin{align}\label{DefNlGWLmax2}
&N^{\text{GW} \Delta}_\ell(L_{\text{max}})=\sqrt{\frac{C_{\ell}^{\Delta}}{2 \ell+1}}\left(\frac{\bar{\Omega}_{\text{GW}}}{4\pi}\right)\left| \mathcal{B}_{\ell m}\right|\,. 
 \end{align}

\section{Results}\label{application}

Before we embark on assessing the impact of cross-correlations on the detectability of the signal, we note that the weight function, $W(r)$, should
be chosen so as to maximize the SNR of cross-correlations.
This can be done as long as radial information (i.e. accurate
redshifts) are available for all galaxies in the survey we cross-
correlate with, which we will assume here. As detailed in \cite{Alonso:2020mva}, the optimal weights can be derived in terms of a
Wiener filter, leading to the result 
\be
W(r)\equiv \frac{ \partial_r \bar{\Omega}_{\text{GW}}(r)}{\bar{\Omega}_{\text{GW}}}\,,
\ee 
Physically this means that we approximately weight all galaxies by a $1/r^2$ factor, hence mimicking the properties of a background mapped in intensity.

\subsection{Signal to noise ratio}

We can now estimate the signal to noise of the auto-correlation and cross-correlation. We assume that shot noise is the dominant noise component, i.e. we assume an ideal experiment. 
The signal to noise of the AGWB auto-correlation is given by (see appendix \ref{SNR} for a derivation)
\be\label{SN11}
\left(\frac{S}{N}\right)_{\text{GW}, \ell}^2 = 2 (2\ell+1)\left(\frac{C_{\ell}^{\text{GW}}}{C_{\ell}^{\text{GW}}+N^{\text{GW, tot}}_\ell}\right)^2\,,
\ee
where $N^{\text{GW, tot}}_\ell=N_{\text{shot}}^{\text{GW}}+N^{\text{GW}}_\ell(L_{\text{max}})$ 
while the one of the cross-correlation is given by 
\begin{align}\label{SN12}
&\left(\frac{S}{N}\right)_{\text{GW}\Delta, \ell}^2 =\\
&= \frac{(2\ell+1)(C_{\ell}^{\text{GW}\Delta})^2}{(C_{\ell}^{\text{GW}\Delta}+N^{\text{GW}\Delta \text{, tot}}_\ell)^2+(C_{\ell}^{\text{GW}}+N^{\text{GW, tot}}_\ell)(C_{\ell}^{\Delta}+N_{\text{shot}}^{\Delta})}\nn\,.
\end{align}
where $N^{\text{GW} \Delta \text{, tot}}_\ell=N_{\text{shot}}^{\text{GW}\Delta}+N^{\text{GW}\Delta}_\ell(L_{\text{max}})$. We recall that $L_{\text{max}}$ corresponds to the angular resolution of the time-residual map. 
One can combine the various multipoles to constrain the global amplitude of the angular power spectrum. Then we define a cumulative SNR as
\begin{align}\label{SN12Cum}
&\left(\frac{S}{N}\right)_{\text{GW}}^2=\sum_{\ell}^{\ell_{\text{max}}} \left(\frac{S}{N}\right)_{\text{GW}, \ell}^2\,,
\end{align}
and similarly for the cross-correlation.  We recall that $\ell_{\text{max}}$ corresponds to the resolution of the energy density map and $\ell_{\text{max}}= 2 L_{\text{max}}$.

In a scenario with $N_f \gg 1$, in  both (\ref{SN11}) and (\ref{SN12}) the dominant contribution to the denominator is coming from the variance of AGWB auto-correlation $N_{\text{shot}}^{\text{GW}}$, which appears quadratically in (\ref{SN11}) and linearly in  (\ref{SN12}). Hence the SNR of cross-correlation will be typically enhanced with respect to the AGWB auto-correlation one. For a finite value of $N_f$, e.g. for $N_f=10$, with $L_{\text{max}} = 20$, the variance \eqref{DefNlGWLmax} of the autocorrelation estimator is much smaller than the corresponding shot noise when no cut to the catalogue is applied. However it becomes slightly larger with a $10^{-15}$ cut and even one order of magnitude larger for the $10^{-16}$ cut as can be checked on the left panel of Fig.\,\ref{Fig1}. When considering the cross-correlation with galaxies, the variance~\eqref{DefNlGWLmax2} is slightly larger than the corresponding shot noise for the full catalogue, and about one order of magnitude larger for the cut catalogues, as can be checked on the right panel of Fig.\,\ref{Fig1}. The denominator of~\eqref{SN12} is nonetheless still dominated by the product $N^{\text{GW, tot}}_\ell C_{\ell}^{\Delta}$, which is enhanced for the cut catalogues by the aforementioned fact that $N^{\text{GW}}_\ell(L_{\text{max}}) > N_{\text{shot}}^{\text{GW}}$.

\subsection{Astrophysical modeling and numerical results}

 We make use of the astrophysical BBH populations of \cite{2013MNRAS.433L...1S,Rosado:2015epa}, and we refer the reader to those papers for full details. In short, the BBH cosmic population is derived from observations of the galaxy mass function $\phi(M_g,z)$ and pair fraction ${\cal F}(z,M_g,q_g)$,  as a function of galaxy mass $M_g$, redshift $z$ and mass ratio of galaxy pairs $q_g<1$. These observables can be combined with a merger time scale $\tau(z,M_g,q_g)$ inferred by numerical detailed simulations to get a galaxy merger rate as: 
\begin{equation}
    \label{eq:galmergerrate}
    \frac{{\rm d}^3n_g}{{\rm d}z{\rm d}M_g{\rm d}q_g} = \frac{\phi(M_g,z)}{M_g {\rm ln}10}\,\frac{{\cal F}(z,M_g,q_g)}{\tau(z,M_g,q_g)} \, \frac{{\rm d}t_r}{{\rm d}z} \, ,
\end{equation}
where $t_r$ is the rest frame cosmic time and $\dd t_r/\dd z$ converts time into redshift for a given cosmology.
The BBH merger rate is then computed by populating galaxies with black holes according to observed scaling relations of the form:
\begin{equation}
    {\rm log}_{10}M_{\rm BH}=\alpha + \beta\, {\rm log}_{10} X \, ,
\end{equation}
where $X$ can be the galaxy bulge mass, or its mid-infrared luminosity or velocity dispersion $\sigma$ (see \cite{2013MNRAS.433L...1S} for a list of those relations).

Here, we take $\phi(M_g,z)$ from \cite{2013ApJ...777...18M}, ${\cal F}(z,M_g,q_g)$ from \cite{2009A&A...498..379D} and $\tau(z,M_g,q_g)$ from \cite{2008MNRAS.391.1489K}. Further, we use the $M-\sigma$ relation given by \cite{2013ARA&A..51..511K}, obtaining a GW background with nominal characteristic strain amplitude at the reference frequency of one year $h_{1\,{\rm yr}}=2.4\times10^{-15}$ where the strain amplitude $h_c(f) = h_{1\,{\rm yr}} (f\times {\rm yr})^{-2/3}$ is related to the GW energy density by $\bar{\Omega}_{\text{GW}}(f)=2 \pi^2/3 (f/H_0)^2 h^2_c(f)$. Using the value $1/H_0 = 14.4\times 10^9\,{\rm yr}$ one finds $\bar{\Omega}_{\text{GW}}(f={\rm yr}^{-1}) \simeq  8\times 10^{-9}$, consistent with the one inferred from the EPTA \texttt{DR2new} analysis \citep{EPTA:2023sfo}. 

Given a BBH merger rate, we compute the cosmic population of inspiraling BBHs, ${\rm d}^3N/({\rm d}z{\rm d}{\cal M}{\rm d}{\rm ln}f)$, by assuming circular-GW driven systems. One can then sample this distribution in the parameter region (${\cal M}>10^{7}$M$_\odot$) $\cap$ ($f>10^{-9}$Hz) $\cap$ ($z<1.3$). One sample draw counts $\approx$100K binaries, with chirp masses, redshift and GW emission frequency, that are used to construct the GW background energy density as described in appendix~\ref{AppC}.\footnote{
A Monte-Carlo sampling of the population returns correctly a discrete number of sources, while the 
semi-analytical result contains spurious contributions
from \emph{fractional sources} that are clearly not present \cite{Sesana:2008mz}. In other words,  sampling is needed because when using an analytical formula, one would also need to add a 'fractions of sources,' while Monte Carlo methods avoid this issue. }

Finally, in order to compute the statistical properties of the GW background, we also need to choose how to distribute the GW sources across the sky. We assume that the distribution of galaxies follows the distribution of dark matter halos with bias factor that we assume to be scale-independent and with redshift evolution given by $b(z)=b_0 \sqrt{1+z}$ and $b_0=1.5$ \cite{WiggleZ:2013kor, Rassat:2008ja}. Finally, as mentioned earlier, we assume that the clustering properties of the massive BBH  population are identical to those of the underlying galaxy distribution. This is, of course, a simplifying approximation, as we expect more massive galaxies to be more likely to host massive black holes. In other words, the population of massive BBH is expected to be \emph{more clustered} than the galaxy distribution. Since refining this bias model would likely strengthen the clustering signal, our simplifying assumption is conservative for the purposes of this study. We also use a constant comoving galaxy density
$a^3 n_{\Gal} \sim 0.1 \,\text{Mpc}^{-3}$.

As already mentioned, we expect the AGWB strain map to be dominated by a few very bright sources: by resolving them, anisotropies of the AGWB generated by the remaining unresolved population are a representative sample of large scale structure anisotropies.  We filter out from the background budget sources whose strain is higher than the current EPTA sensitivity curve \cite{InternationalPulsarTimingArray:2023mzf}, roughly corresponding to $h_{\text{eff}}>10^{-14}$ around the sweet spot of the sensitivity curve. We have here introduced an effective strain $h_{\rm eff}$ defined by 
\be\label{eq:h2g2}
h_{\rm eff}^2(f) = h^2(f) f T_{\rm obs}\,,
\ee
where $T_{\rm obs}=16\,{\rm yr}$ and $h(f)= 8/(\sqrt{10} D_L) G \M_z  [\pi G \M _zf]^{2/3}$ with $\mathcal{M}_z=(1+z)\mathcal{M}$ is the polarization and inclination averaged strain (see e.g.~\cite{Sesana:2008mz} with proper factors of redshift accounted for).  As the sensitivity of a given PTA observatory will improve, more and more sources will be detectable individually: when deriving forecasts for future experiments, the filtering should be performed following the iterative procedure of \cite{2024arXiv240712078T}, 
accounting in the computation of the expected noise budget for both a pure noise contribution and contribution of the unresolvable background component.\footnote{We stress that both these contributions are accounted for when computing the sensitivity curve using  real PTA data.} Since our goal is just to quantify the effect of filtering without targeting a specific future PTA mission, 
we test the effect of moving the threshold on $h_{\rm eff}$, 
filtering out sources with $h_{\text{eff}}>10^{-15}$ and with $h_{\text{eff}}>10^{-16}$, the latter corresponding to the plateau of the SKA sensitivity curve \cite{2024arXiv240712078T}.\footnote{We stress again that in the context of SKA a proper filtering procedure would need to account not only for the instrumental noise component, but also for the background budget. In this respect, a threshold $h_{\text{eff}}>10^{-16}$ for filtering out resolvable events with SKA is optimistic.}
We refer to these three values for the filtering threshold as "no cut" (as the EPTA sensitivity effectively does not allow to resolve any individual source), "$10^{-15}$ cut" and "$10^{-16}$ cut".

\begin{figure}
\centering
\includegraphics[width=\columnwidth]{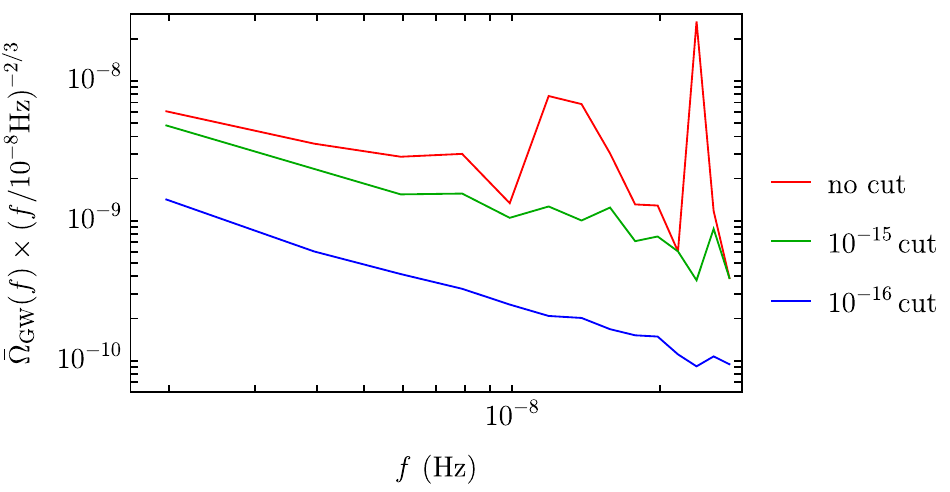}
\caption{Background energy density  $\bar{\Omega}_{\rm GW}(f)$ as function of frequency, for the three different cut-off values used in the filtering procedure. }
\label{Fig0}
\end{figure}

In Fig.\,\ref{Fig0} we show the background energy density as function of frequency, for the three aforementioned cut-off values. In the "no cut" situation, after averaging over modulations, the scaling is roughly $\propto f^{2/3}$ as expected. The bumps in the red curve are due to the contribution of a few very bright sources that dominate the total signal at high frequencies (plotting source contributions individually, they would appear as spikes at a given frequency). We observe that the filtering procedure has the effect of eliminating these high-frequency bumps and changing the low-frequency scaling. Indeed for a given threshold $h_{\rm eff}^{\rm cut}$, we infer from~\eqref{eq:h2g2} that $\M_{\rm max} \propto (h_{\rm eff}^{\rm cut})^{3/5} f^{-7/10}$, hence the larger the frequency, the more sources are removed from the catalog, leaving behind a smaller residual amplitude. 

\begin{figure*}
\centering
\includegraphics[width=0.83\columnwidth]{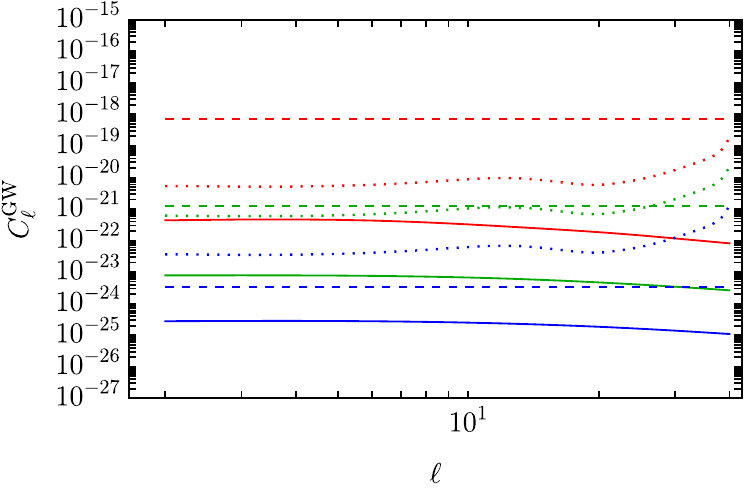}\,\,
\includegraphics[width=1.17\columnwidth]{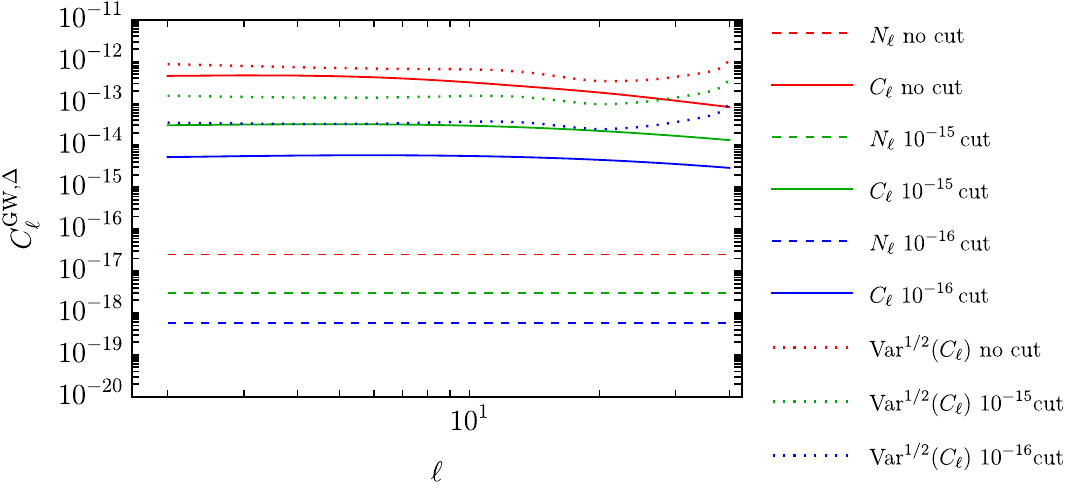}
\caption{Angular power spectrum of auto-correlation (left panel) and cross-correlation (right panel) and corresponding shot noise and finite resolution map noise contributions. The signal is integrated over the PTA frequency range. We are here using  $L_{\text{max}} = 20$ and $N_f=10$. }
\label{Fig1}
\end{figure*}

\begin{figure*}
\centering
\includegraphics[width=0.82\columnwidth]{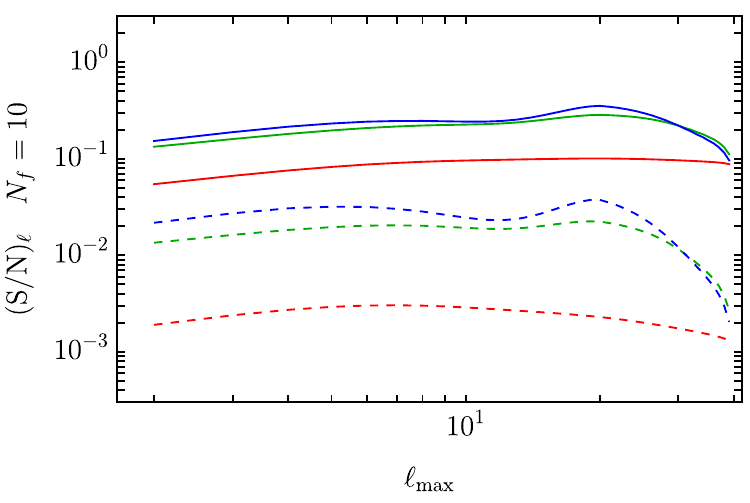}\,\,
\includegraphics[width=1.18\columnwidth]{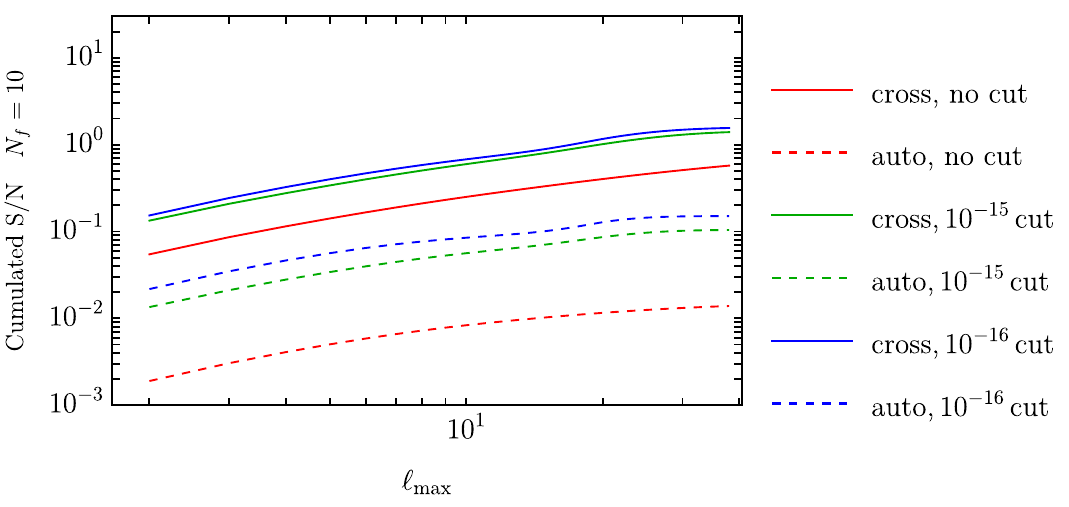}\\
\includegraphics[width=0.82\columnwidth]{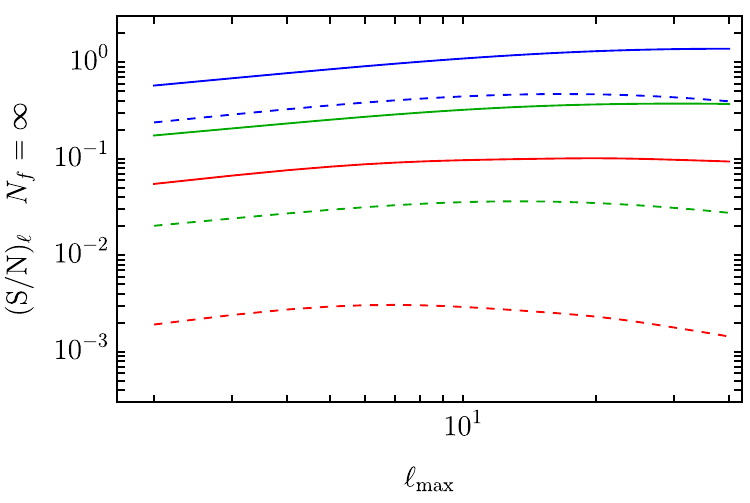}\,\,
\includegraphics[width=1.18\columnwidth]{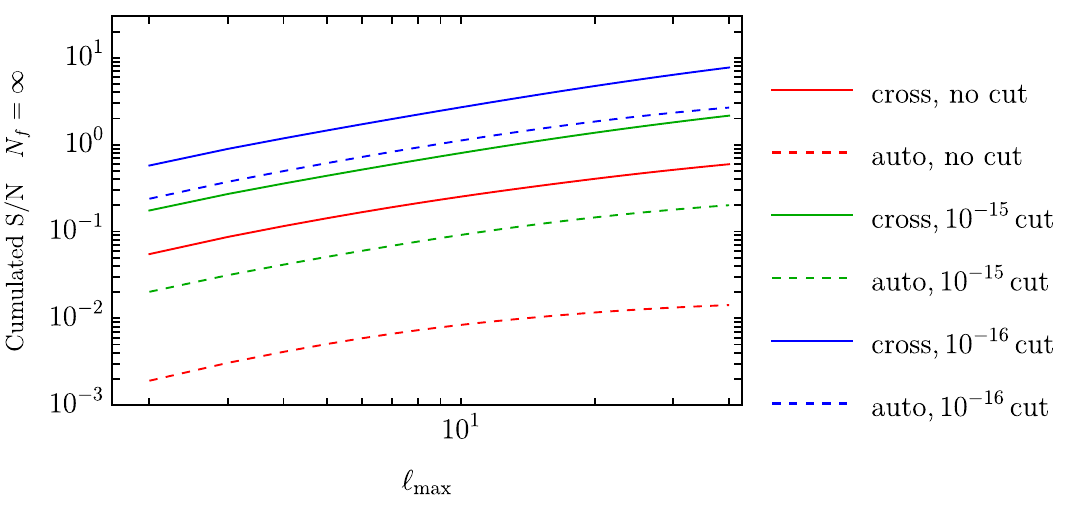}
\caption{\emph{Left}: SNR as function of the multipole $\ell$ of cross-correlation (continuous lines) and of auto-correlation (dashed lines) for $N_f=10$ and for an ideal experiment in which the shot noise is the only noise component. \emph{Right}: corresponding cumulative SNR as function of $\ell_{\text{max}}$. The first line of panels uses $N_f=10$ whereas the second line is idealized with $N_f = \infty$. Both cases consider an angular resolution set by $L_{\text{max}}=20$.}
\label{Fig2}
\end{figure*}

Fig.\,\ref{Fig1} represents for the three situations the angular power spectrum of auto- and cross-correlations, and the corresponding shot noise contributions.\footnote{Our results for the cross-spectrum are compatible in order of magnitude with those of \cite{Sah:2024etc} (Fig.\,3), where a different set of astrophysical models is used.} We see that while the auto-correlation map is shot noise dominated ($N^{\rm GW}_{\rm shot} \gg C_\ell^{\rm GW}$), the shot noise contribution to the cross-correlation is much suppressed ($N^{{\rm GW}\Delta}_{\rm shot} \ll C_\ell^{{\rm GW}\Delta}$). The finite resolution noise as well is much suppressed for the cross-correlation than for the auto-correlation.  
We observe that the filtering procedure reduces both the angular power spectrum of the clustering signal and the  noise components, for both auto- and cross-angular power spectra.

In Fig.\,\ref{Fig2} (left panel) we show the SNR for auto- and cross-correlations (for the three cut-off values) as a function of the multipole $\ell$. On the right panel we show the corresponding cumulative SNR, as a function of $\ell_{\text{max}}$. We consider both a scenario with $N_f=10$ and an ideal scenario where shot noise is the only noise component (bottom plots).  As expected, the SNR of the cross-correlation outperforms the auto-correlation one (by more almost two orders of magnitudes in the absence of a cut-off). This is due to the aforementioned hierarchy between signal and noise components along with $N^{\Delta}_{\rm shot} \ll C_\ell^\Delta$, i.e. the fact that the auto-correlation SNR scales as $\sqrt{2 (2\ell+1)} C_{\ell}^{\text{GW}}/N^{\text{GW, tot}}_\ell$ whereas for cross-correlation it scales as $\sqrt{(2\ell+1)} C_{\ell}^{\text{GW}\Delta}/\sqrt{ C_{\ell}^{\Delta} \, N^{\text{GW, tot}}_\ell } $.

We observe that even in the ideal case where the instrumental noise is set to zero (such that $N^{\text{GW, tot}}_\ell = N_{\text{shot}}^{\text{GW}}$), the SNR of single multipoles is never much larger than 1, unless a very low threshold of detection of single events is chosen. However, the cumulative SNR is above 1 already at $\ell_{\text{max}}=4$ (corresponding to $\sim 25$ pulsars), for a filtering threshold of $10^{-16}$, and it is above 1 for $\ell_{\text{max}}$=13 for a threshold of $10^{-15}$. This indicates that, while the reconstruction of the shape of the angular power spectrum will be challenging, it will be potentially possible to constrain the global amplitude of the spectrum, by combining the information from different multipoles. For a realistic experiment (i.e. $N_f$ finite) the situation stays qualitatively the same: when $N_f=10$, detectability can only be achieved using multipoles up to $\ell_{\text{max}}=30$, with a threshold of $10^{-15}$.

We conclude observing that to obtain these results, we have fixed the value of the comoving galaxy number density equal to $a^3 n_\text{G}=0.1$ Mpc$^{-3}$, however our results for the SNR depend very mildly on this choice.  Indeed the SNR of the auto-correlation map does not depend on the galaxy number density assumed, while, for any reasonable value of $a^3 n_\text{G}$, the dominant contribution to the noise of the cross-correlation map comes from the $\propto N_{\ell}^{\text{GW, tot}}C_{\ell}^{\Delta}$ contribution in the denominator of eq.\,(\ref{SN11}), which does not depend on the galaxy number density.

\section{Discussion and conclusions}

In this article, we explored the potential of cross-correlation between a background map and a dense galaxy map to obtain a first detection of background anisotropies. We considered both the case of a perfect instrument (for which pulsar white noise is negligible), where shot noise is the dominant source of noise, and the case of a realistic survey in which an additional noise component arises due to the finite resolution of a time residual map (finite resolution noise). We derived from first principles the expressions of shot noise and time resolution noise for both the auto- and cross-correlation spectrum. The magnitude of the finite resolution noise is  set by the effective number of frequencies, which in turn depends on the number of pulsar monitored, pulsar white noise level and observation time, as discussed in detail in \cite{Pitrou:2024scp}. For a perfect experiment (pulsar white noise is negligible) one has  $N_f=2 T f_{\text{max}} \gg 1$, while for a realistic survey we consider $N_f=10$.   

Using catalogs of massive BBH from \cite{2013MNRAS.433L...1S,Rosado:2015epa}, we showed that, unlike the auto-correlation, the cross-correlation map is not shot noise dominated: for an ideal experiment, the SNR of the cross-correlation outperforms the auto-correlation by almost two orders of magnitude.  We also tested the effect of a filtering procedure that removes the brightest sources from the total background: this procedure further reduces shot noise, as the brightest sources dominate the background budget, obscuring the contribution of the unresolved population that traces the large-scale structure. A similar result was recently found in \cite{Semenzato:2024mtn}, relying on numerical simulations. However, while our findings are consistent with the simulation results of \cite{Semenzato:2024mtn}, our analytic treatment disagrees with the explanation proposed in that reference.\footnote{Through a detailed analytic derivation, we demonstrate that shot noise is, in fact, never fully eliminated in cross-correlation maps, contrary to what is claimed in eq. (7) of \cite{Semenzato:2024mtn}, which states that the cross-correlation map is shot noise free.}  We show that the finite resolution noise is as well considerably reduced in the cross-correlation.    

For both the cases of  a perfect experiment and a realistic one, when filtering out bright sources, we had to assume a threshold value
to determine whether a source is resolvable individually. To
do so, we considered the current EPTA sensitivity curve (effectively resulting in no resolvable sources), a threshold value
corresponding to SKA sensitivity, and an intermediate scenario. We found that even in the ideal case where the instrumental noise is set to zero (the noise is just shot noise), the SNR of single multipoles is never much larger than 1, unless a very low threshold of detection of single events is chosen. However, the cumulative SNR is above 1 already at $\ell_{\text{max}}=4$ (corresponding to $\sim 25$ pulsars), for a filtering threshold of $10^{-16}$, and it is above 1 for $\ell_{\text{max}}$=13 for a threshold of $10^{-15}$. This indicates that, while the reconstruction of the shape of the angular power spectrum will be challenging, it will be potentially possible to constrain the global amplitude of the spectrum, by combining the information from different multipoles. For a realistic experiment (i.e. $N_f$ finite) the situation stays qualitatively the same: when $N_f=10$, detectability can be achieved using multipoles up to $\ell_{\text{max}}=30$, with a threshold of $10^{-15}$.  We stress again that the case of $N_f=10$ is rather futuristic: as shown in detail in \cite{Pitrou:2024scp} the number of effective frequencies $N_f$ is in fact a decreasing function of the map multipole $L$. Assuming the pulsar noise characteristics of~\cite{NANOGrav:2023hde}, getting $N_f(L=5)\sim 1$  would require 450 pulsars monitored during
25 years. Of course the situation would improve significantly when pulsar white noise is reduced with respect to the model of \cite{NANOGrav:2023hde}, e.g. by increasing the cadence of pulsar observations. A future study will be dedicated to deriving more realistic forecasts for future PTA experiments and galaxy surveys such as Euclid, properly factoring in the effects of white noise reduction in future PTA surveys and realistic filtering procedures.

\section*{Acknowledgement} 

The work of GC and CP is supported by CNRS. The work of GC and MP has received financial support by the SNSF Ambizione grant \emph{Gravitational wave propagation in the clustered universe}. We are very grateful to Marc Kamionkowski for having pointed out that finite resolution noise was missing in our final results. We thank Camille Bonvin and Nastassia Grimm for interesting discussions at different stages of this work. 
AS acknowledges financial support by the ERC CoG "B Massive" (Grant Agreement: 818691) and ERC AdG "PINGU" (Grant Agreement: 101142079).

\newpage

\appendix

\section{General properties of compound statistics}\label{Appcompound}

Let us consider $N$ being a random variable with variance $\langle N^2\rangle-\langle N\rangle^2$ and mean $\langle N\rangle$. Then 
for a generic function $f(N)$
\be
\langle f(N) \rangle = \int f(N) p(N) \dd N\,.
\ee
We are interested in the statistics of the compound random variable $Y$ given by 
\be
Y=\sum_{i=1}^N X_i\,,
\ee
where $X_i$ is either $1$ or $0$, i.e. it is a binomial variable $X_i\sim B(1, \beta)$.  It has a probability $p(Y)$, such
that
\be
p(Y) = \int p(Y | N) p(N)\dd N\,.
\ee
The conditional probability $p(Y|N)$ can be found from the fact
that it is the sum of $N$ random variables (it is a convolution of
probabilities), but we do not need to go to this level of
sophistication to get convinced that
\be\label{A5}
\langle Y\rangle_{N} = \int Y p(Y|N) \dd Y = N \langle X \rangle \,.
\ee
The subindex on the left hand side means that is an average at $N$ being fixed. For the second moment one has
\be
\langle Y^2\rangle_{N} = \int Y^2 p(Y|N) \dd Y = N \langle X^2\rangle +
N(N-1)\langle X \rangle^2\,,
\ee
which is obtained 
 by separation of the $X_i^2$ and the $X_i X_j$ with $i\neq j$.

We can work out the first moments of $Y$ from $p(Y)$. For a generic function of $Y$, we have
\bea
\langle f(Y) \rangle &=& \int f(Y) p(Y) \dd Y \\
&=& \int\int  f(Y) p(Y|N) p(N) \dd N \dd Y\nonumber\\
&=& \int p(N) \dd N \left(\int f(Y) p(Y|N) \dd Y\right)\nonumber\\
&=& \int \langle f(Y) \rangle_N\, p(N) \dd N\,,\nonumber
\eea
where in the last line  the remaining randomness of $N$ has to be treated. The average $\langle Y \rangle$ can be easily computed 
\be
\langle Y \rangle = \int \langle Y \rangle_N p(N) \dd N =\int \langle X \rangle  N p(N) \dd N =\langle N \rangle \langle X \rangle\,.
\ee
Similarly for the second moment
\bea
\langle Y^2 \rangle &=& \int \langle Y^2 \rangle_N p(N) \dd N \nonumber\\
&=& \int \left(N \langle X^2 \rangle + N(N-1) \langle X\rangle^2\right)p(N)
\dd N \nonumber\\
&=& \langle N \rangle (\langle X^2 \rangle-\langle X\rangle^2) + \langle N^2 \rangle \langle X \rangle^2\,.
\eea
Hence we obtain the variance of $Y$
\begin{align}
\langle Y^2 \rangle - \langle Y\rangle^2 &= \langle N \rangle (\langle
X^2 \rangle-\langle X\rangle^2) + (\langle N^2 \rangle -\langle
N\rangle^2)\langle X \rangle^2\nn\\
&=\langle N \rangle \langle X^2 \rangle\,,\
\label{EqNX2}
\end{align}
where the last equality holds when $N$ is a Poisson variable. 
This derivation is valid whatever the statistics for $X$, and  $\langle X^2 \rangle = \langle X \rangle$ if $X\sim B(1, \beta)$. Following the same kind of reasoning we can also get cross
correlations between $Y$ and $N$, and
\bea
\langle Y N \rangle &=& \int \langle Y \rangle_N  \, N p(N) \dd N \nonumber\\
&=&\int \langle X \rangle  N^2 p(N) \dd N \nonumber\\
&=& \langle X \rangle \langle N^2\rangle\nn\\
&=& \langle X \rangle (\langle N\rangle + \langle N\rangle^2)\,,
\eea
where the last equality holds when $N$ is Poisson variable. 
Then 
\be
\langle Y N \rangle -  \langle Y\rangle \langle N \rangle =\langle X
\rangle \text{Var}(N)=\langle X\rangle \langle N \rangle\,,\label{EqYN}
\ee
where the last equality holds when $N$ is a Poisson variable.

\section{Derivation of the finite resolution noise}\label{NewApp}

This appendix is a summary of the results of \cite{Pitrou:2024scp} on anisotropic estimators for a stochastic background. The goal is to derive the variance affecting our ability to construct the angular power spectrum of the background energy density, due to the finite resolution of the time-residual map. 

The statistics of the multipoles $\hat{z}_{L M}(f)$ of pulsar redshifts  is given by 
\be\label{Correlationlmf}
\langle \hat{z}_{L_1 M_1}(f_1) \hat{z}^{\star}_{L_2
  M_2}(f_2)\rangle_{\AmpGW} = \delta_{L_1 L_2} \delta_{M_1 M_2}\delta(f_1-f_2)
C^{z}_{L_1}(f_1)\,,
\ee
where the average $\langle \dots \rangle_{\AmpGW}$ is over the GW realisation and 
\be
C^{z}_L(f) = \hat{\cal P}(f) C_L^{\rm HD}\,,
\ee
where $C_L^{\rm HD}$ is the multipole decomposition of the Hellings-Downs curve 
while $\hat{\cal P}(f)$ is the (two-sided, meaning that needs to be integrated on both positive and negative frequencies) spectral function, related to the gravitons distribution function as 
\be\label{DefPf}
\hat{\cal P}(f) \equiv 4\pi f^2 {\AmpGW}(2 \pi f)\,.
\ee
The isotropic part of the spectral density is related to the background energy density, which can be quantified in units of the critical energy density with 
\be\label{OP}
\bar{\Omega}_{\rm GW}(f) = \frac{8 \pi^2}{3 H_0^2} f^3  \hat{\cal P}(f)=\frac{32 \pi^3}{3 H_0^2} f^5  \mathcal{A}(2 \pi f)\,.
\ee

In full generality, the graviton distribution function can depend on directions. Let us parameterize its angular dependence as
\be\label{AnisoStat}
\AmpGW(\gr{k}) = \AmpGW(k) \hat{\AmpGW}(\hat{\gr{k}})\,,\quad \hat{\AmpGW}(\hat{\gr{k}}) = \sum_{\ell m} \hat{\AmpGW}_{\ell m} Y_\ell^m(\hat{\gr{k}})\,,
\ee
where $k=|\gr{k}|$ and $\hat{\AmpGW}_{\ell\,-m}=(-1)^m\hat{\AmpGW}_{\ell m}^\star$  
since the GW spectrum is real valued. This decomposition assumes the existence of a factorization between a spectral shape $\AmpGW(k)$, and an angular-dependent function  $\hat{\AmpGW}(\hat{\gr{k}})$. 
We get from (\ref{Splitbackgroundperturbations}) and (\ref{OP})
\be\label{deltaOP}
\left(\delta\Omega_{\rm GW}\right)_{\ell m} =\frac{8 \pi^2}{3 H_0^2} f^5  \mathcal{A}(2 \pi f)\hat{\mathcal{A}}_{\ell m}\,. 
\ee

Let us discuss the effects of an anisotropic spectrum: if $\hat{\AmpGW}(\hat{\gr{k}}) \neq 1$, the correlation in equation~\eqref{Correlationlmf} acquires off-diagonal contributions, and one obtains 
\cite{Pitrou:2024scp}
\begin{align}\label{GeneralCorrelation}
&\langle \hat{z}_{L_1 M_1}(f_1) \hat{z}_{L_2 M_2}(f_2) \rangle_{\AmpGW} = \delta(f_1 + f_2)\hat{\cal P}(f_1)F_{L_1} F_{L_2}\\
&\times\sum_{\ell m}\delta^{\rm even}_{L_1+L_2+\ell} \frac{\hat{\AmpGW}^\star_{\ell m}}{\sqrt{4\pi}}  \sqrt{2\ell+1}\troisj{\ell}{L_1}{L_2}{m}{M_1}{M_2}\troisj{\ell}{L_1}{L_2}{0}{-2}{2}\,,\nonumber
\end{align}
where 
the quantities $F_{L}$ are defined as 
\be\label{Fellcompact}
F_L\equiv \sqrt{4\pi (2L+1)}\sqrt{\frac{(L-2)!}{(L+2)!}}\,,
\ee
and $\delta^{\rm even}_{L_1+L_2+\ell}=1$ if $L_1+L_2+\ell$ is even and vanishes if it is odd. The isotropic result~\eqref{Correlationlmf} is easily recovered when using $\hat{\AmpGW}_{\ell m}=\delta_\ell^0 \delta_m^0 \sqrt{4 \pi}$. The HD correlation is a signature of an isotropic spectral function $\AmpGW(k)$. The anisotropic part $\hat{\AmpGW}(\hat{\gr{k}})$ of the GW background is reflected instead in the off-diagonal correlations in the $(L, M)$ space. 

Our goal is to isolate the ${\AmpGW}_{\ell m}$ modes and this has to be done by exploiting the orthonormality of Wigner 3j. Hence we introduce the combinations (contractions of the 2-point function with Wigner 3j symbol)
\begin{align}\label{DefElllLM}
{\cal E}^{\ell m}_{L_1 L_2}(f_1,f_2) &\equiv \sqrt{2\ell+1}\sum_{M_1 M_2}\troisj{\ell}{L_1}{L_2}{m}{M_1}{M_2} \nonumber\\
&\qquad\times \hat{z}_{L_1 M_1}(f_1)\hat{z}_{L_2 M_2}(f_2)\,,
\end{align}
whose average over the GW realisation can be computed using~\eqref{GeneralCorrelation} and using properties of Wigner symbols, and is given by 
\begin{align}\label{EqAverageELMl1l2}
\langle {\cal E}^{\ell m}_{L_1 L_2}(f_1, f_2) \rangle_\AmpGW &= \delta_{L_1+L_2+\ell}^{\rm even}\delta(f_1+f_2) \hat{{\cal P}}(f_1)\nonumber\\
&\quad\frac{\hat{\AmpGW}^\star_{\ell m}}{\sqrt{4\pi}} F_{L_1} F_{L_2} \troisj{\ell}{L_1}{L_2}{0}{-2}{2}\,.
\end{align}

Hereafter we consider only parameters such that $L_1+L_2+\ell$ is even. In the case where the anisotropy is mild, we can assume the covariance is only due to the isotropic part. In that case,  the covariance takes the simple form 
\begin{align}\label{VarE}
&\text{Var}({\cal E}^{\ell m}_{L_1 L_2})\equiv\langle{\cal E}^{\ell m}_{L_1 L_2}(f_1,f_2) {\cal E}^{\ell'm' \star}_{L'_1 L'_2}(f'_1,f'_2)\rangle_\AmpGW\\
&-\langle{\cal E}^{\ell m}_{L_1 L_2}(f_1,f_2)\rangle_\AmpGW \langle {\cal E}^{\ell'm' \star}_{L'_1 L'_2}(f'_1,f'_2)\rangle_\AmpGW\nonumber\\
&\simeq\delta_\ell^{\ell'}\delta_m^{m'}\left[\delta_{L_1}^{L'_1}\delta_{L_2}^{L'_2} \delta(f_1-f_1')C^z_{L_1}(f_1) \delta(f_2-f_2')C^z_{L_2}(f_2)\right. \nonumber\\
&\qquad\qquad+\left.\delta_{L_1}^{L'_2}\delta_{L_2}^{L'_1}\delta(f_1-f_2')C^z_{L_1}(f_1)\delta(f_2-f_1')C^z_{L_2}(f_2)\right]\,.\nonumber
\end{align}
If we want to combine the expressions~\eqref{DefElllLM} to form an optimal estimator for the $\hat{\AmpGW}_{\ell m}$, we must normalize them and weight them by their inverse variance~\cite{Hotinli:2019tpc}. Explicitly, after having discretized the frequency space with bins of width $1/T$, one finds the following estimator
\begin{align}
&\left(\hat{\AmpGW}^{L_1 L_2}_{\ell m, f}\right)^e \label{EqdefAe} \\
&=\frac{\sqrt{4\pi}}{T F_{L_1}F_{L_2}}\frac{1}{\hat{\mathcal{P}}(f)}\troisj{\ell}{L_1}{L_2}{0}{-2}{2}^{-1}{\cal E}^{\ell m}_{L_1 L_2}(f,-f)^{*}\nonumber\,,
\end{align}
where on the left hand side we have explicitly indicated that one can build an estimator for each given value of $L_1$ and $L_2$. From~\eqref{EqAverageELMl1l2} the mean of this object is trivially given by $\hat{\AmpGW}^{L_1 L_2}_{\ell m}$ as we considered only cases for which $L_1+L_2+\ell$ is even. 
Assuming that anisotropies are mild, and using (\ref{VarE}), the variance of this object for $f \neq 0$ is given by 
\be
\text{Var}\left(\hat{\AmpGW}^{L_1 L_2}_{\ell m, f}\right)^e=\frac{4\pi}{(2L_1+1)(2L_2+1)}\troisj{\ell}{L_1}{L_2}{0}{-2}{2}^{-2}\,. \label{EqVarAe}
\ee  
We then combine all the estimators $\left(\hat{\AmpGW}^{L_1 L_2}_{\ell m, f}\right)^e$ with inverse variance weighting to obtain the minimum-variance estimator. However since
\be\label{EqSymmetryALM}
\left(\hat{\AmpGW}^{L_1 L_2}_{\ell m, f}\right)^e = \left(\hat{\AmpGW}^{L_2 L_1}_{\ell m, -f}\right)^e\,,
\ee
we restrict to $f>0$ while considering all $\ell_1,\ell_2$ combinations so as to avoid double counting. The estimator is therefore
\be
\hat{\AmpGW}_{\ell m}^e=\frac{\sum_{L_1 L_2}\sum_{f>0} \delta^{\text{even}}_{L_1+L_2+\ell} \left(\hat{\AmpGW}^{L_1 L_2}_{\ell m, f}\right)^e \text{Var}^{-1}\left(\hat{\AmpGW}^{L_1 L_2}_{\ell m, f}\right)}{\sum_{L_1 L_2}\sum_{f>0} \delta^{\text{even}}_{L_1+L_2+\ell} \text{Var}^{-1}\left(\hat{\AmpGW}^{L_1 L_2}_{\ell m, f}\right)}\,,
\ee
The variance of this estimator is computed from 
\be
\text{Var}^{-1}\left(\hat{\AmpGW}^e_{\ell m}\right)=\sum_{L_1 L_2}\sum_{f>0} \delta^{\text{even}}_{L_1+L_2+\ell} \text{Var}^{-1}\left(\hat{\AmpGW}^{L_1 L_2}_{\ell m}\right)^e\,,
\ee
and we get
\begin{align}\label{varA}
&\text{Var}^{-1}\left(\hat{\AmpGW}^e_{\ell m}\right)\\
&=\frac{1}{8\pi}\sum_{L_1 L_2 }N_f\delta_{L_1+L_2+\ell}^{\text{even}}(2L_1+1)(2L_2+1) \troisj{\ell}{L_1}{L_2}{0}{-2}{2}^2\nonumber\,,
\end{align}
where $N_f\equiv \text{min}\left(N_f(L_1), N_f(L_2)\right)$ and 
the sum is over all values of $L_1$ and $L_2$. 
In this equation we assumed that in the absence of instrumental  noise all frequencies equally contribute, and $N_f$ is equal to the number of frequency modes, i.e. $N_f = 2 T f_{\rm max}$ for a perfect experiment, so that $N_f/2$ is the number of positive frequency modes. In the presence of noise, the effective number of frequency modes has to be evaluated following the procedure described in section VII C2 of \cite{Pitrou:2024scp}, and is a function of the map multipole $L$, which depends on the pulsar noise level (beside pulsar number and observation time), see \cite{Pitrou:2024scp} for a detailed discussion. Hence here $N_f$ represents an effective quantity, which we introduce for practical purposes, even if a realistic evaluation for a given instrument should keep the dependence on $L_1$ and $L_2$. This result (\ref{varA}) agrees with what obtained in \cite{Hotinli:2019tpc} (where it is implicitly assumed that $N_f=1$). 
 Notice that all these sums run over $0\leq \ell \leq 2 L_{\text{max}}$ and  $L_1, L_2 \leq L_{\text{max}}$ with $L_{\text{max}}\sim \sqrt{N_p}$, and $N_p$ is the number of pulsars in the network. 

 We conclude this section by relating this anisotropy estimator to the estimator of the energy density angular power spectrum, which is usually the observables used to quantify anisotropies \cite{Cusin:2017fwz}. Using (\ref{DefPf}), (\ref{OP}) and \eqref{Splitbackgroundperturbations} we build the estimator for the multipoles defined in~\eqref{Eq:Sec2Sph}
 \begin{align}
a^e_{\ell m}&=\frac{8 \pi^2}{3 H_0^2} f^5 \AmpGW(2\pi f) \hat{\AmpGW}_{\ell m}^e\,,
 \end{align}
 whose mean is $a_{\ell m}$ by construction, and whose variance can be trivially computed from the one of $\hat{\AmpGW}_{\ell m}^e$. 
 An estimator for the angular power spectrum of this quantity is given by
 \be
C_\ell^{\text{GW}e}=\frac{1}{2\ell+1}\sum_m a^e_{\ell m} a^{e*}_{\ell m}\,,
\ee
 with variance given by 
 \begin{align}
&\text{Var}\left(C_\ell^{\text{GW}e}\right)=\frac{2}{(2 \ell+1)^2}\\
&\times \sum_m \left[\text{Var}^4\left(a_{\ell m}^e\right) + 2 \text{Var}\left(a_{\ell m}^e \right) a_{\ell m} a^\star_{\ell m} \right]\,.\nonumber
 \end{align}
We stress again that the average is an average over GW realizations, and there is a residual stochasticity due to cosmological initial conditions  (see also \cite{Grimm:2024hgi} for a detailed discussion of this double average procedure). In other words, the result above holds for any cosmological realization.  
To obtain predictions, one has to take an additional average over cosmological initial conditions and we get 
\begin{align}
\label{VarCl}&\text{Var}\left(C_\ell^{\text{GW}e}\right)=\frac{2}{2 \ell+1}\nonumber\\
&\left[\left(\frac{\bar{\Omega}_{\text{GW}}}{4\pi}\right)^4  \text{Var}^4 \left(\hat{\AmpGW}_{\ell m}^e\right)+2  \text{Var}\left(\hat{\AmpGW}_{\ell m}^e\right)\left(\frac{\bar{\Omega}_{\text{GW}}}{4\pi}\right)^2  C_\ell^{\text{GW}}\right]\,,
 \end{align}
 which can be directly computed from (\ref{varA}). It follows that for a given pulsar number $N_p$ (determining the angular resolution of a time-residual map, as $L_{\text{max}}\sim \sqrt{N_p}$), (\ref{VarCl}) provides us with the precision with which, in a perfect experiment, the angular power spectrum of the background energy density can be computed from a time-residual map.

 We now consider the cross-correlation with the galaxy number counts, under the assumption that the galaxy map has an infinite angular resolution. We compute 
  \be
C_\ell^{\text{GW}\Delta e}=\frac{1}{2\ell+1}\sum_m a^e_{\ell m} b^{*}_{\ell m}\,,
\ee
where $b_{\ell m}$ are the multipoles of the galaxy number counts, defined in (\ref{blm}). Repeating similar steps as above we get 
 \begin{align}
\label{VarClcross}&\text{Var}\left(C_\ell^{\text{GW} \Delta e}\right)=\frac{C_\ell^{\Delta} }{2 \ell+1}\left(\frac{\bar{\Omega}_{\text{GW}}}{4\pi}\right)^2  \text{Var} \left(\hat{\AmpGW}_{\ell m}^e\right) \,,
 \end{align}
 which can be directly computed from (\ref{varA}).

\section{Signal to noise derivation}\label{SNR}

Let us assume that our observables are the angular power spectra of the AGWB  and galaxy auto-correlations and their cross-correlation angular spectrum, i.e.  $C_{\ell}^{\text{GW}}$, $C_{\ell}^{\Delta}$ and $C_{\ell}^{\text{GW}\Delta}$ are the observables. The estimator (full sky) of the angular power spectrum is
\be\label{hatCab}
\hat{C}_{\ell}^{ab}=\frac{1}{2\ell+1}\sum_m a_{\ell m}^a (a_{\ell m}^b)^*\,, 
\ee
where $a_{\ell m}^{a, b}$ are Gaussian maps and the indices are defined as $a=(\text{GW}, \Delta)$ and $b=(\text{GW}, \Delta)$.   
We compute the covariance of this estimator. 
To make the notation compact, we  introduce $X=(C_{\ell_{\text{min}}}^{\text{GW}}, C_{\ell_{\text{min}}}^{\Delta}, C_{\ell_{\text{min}}}^{\text{GW}\Delta}, ..., C_{\ell_{\text{max}}}^{\text{GW}}, C_{\ell_{\text{max}}}^{\Delta}, C_{\ell_{\text{max}}}^{\text{GW}\Delta})$. Then the associated covariance matrix is block diagonal 
\be
C=\left(
\begin{array}{ccc}
B_{\ell_{\text{min}}}&&0\\
&...&\\
0&&B_{\ell_{\text{max}}}\\
\end{array}
\right)\,,\nonumber
\ee
with 3$\times$3 blocks 
\begin{align}
&B_{\ell}=\frac{1}{2\ell+1} \times \nn\\
&\left(
\begin{array}{ccc}
2(C_{\ell}^{\text{GW}})^2&2(C_{\ell}^{\text{GW}\Delta})^2&2(C_{\ell}^{\text{GW}})(C_{\ell}^{\text{GW}\Delta})\\
2(C_{\ell}^{\text{GW}\Delta})^2&2(C_{\ell}^{\Delta})^2&2(C_{\ell}^{\Delta})(C_{\ell}^{\text{GW}\Delta})\\
2(C_{\ell}^{\text{GW}})(C_{\ell}^{\text{GW}\Delta})&2(C_{\ell}^{\Delta})(C_{\ell}^{\text{GW}\Delta})&C_{\ell}^{\text{GW}}C_{\ell}^{\Delta}+(C_{\ell}^{\text{GW}\Delta})^2
\end{array}
\right)\,,\nonumber
\end{align}
where  it is understood that each entry includes a noise part, i.e. one has to replace $C_{\ell}^{ab}\rightarrow C_{\ell}^{ab}+N_\ell^{ab}$. 
We have all the ingredients to compute the (non-Gaussian) likelihood of the angular power spectra $X$ and compute the corresponding Fisher matrix deriving it with respect to the parameters $\theta_a$. 
The Fisher matrix is defined as 
\be
F_{ab}=\Big<-\frac{\partial^2 \ln\mathcal{L}}{\partial\theta_a\partial\theta_b}\Big>\,, 
\ee
where $\theta_a$ are a set of parameters of the model. It is possible to show that only the Gaussian part of the likelihood enters the Fisher matrix. We want to estimate the signal to noise of the galaxy map, hence we parametrize (adding a generic noise component to the angular power spectrum of auto- and cross-correlations)  
\be
X_{\ell}=(\mathcal{C}^2 C_{\ell}^{\text{GW}}+N_\ell^{\text{GW}}, C_{\ell}^{\Delta}+ N_\ell^{\Delta}, \mathcal{C}  C_{\ell}^{\text{GW}\Delta}+N_\ell^{\text{GW}\Delta})\,,
\ee
where $\mathcal{C}$ is a book keeping parameter for the amplitude. 
Then
\be\label{EqGeneralFAA}
\left(F_{\mathcal{C}}\right)_{\ell}=\left(\frac{S}{N}\right)_{\ell}^2=(\partial_{\mathcal{C}}
X_\ell) \cdot  B_{\ell}^{-1} \cdot (\partial_{\mathcal{C}} X_{\ell})\,,
\ee
evaluated in ${\mathcal{C}}=1$.
If we use only the information from the auto-correlation, we get~\eqref{SN11} 
while using only cross-correlation we obtain~\eqref{SN12}.

\section{The GW energy density}\label{AppC}

Let us introduce some basic quantities that we extract from the catalog. The chirp mass of a source  $\mathcal{M}$ is given by
\be
\label{eq:chirp}
\mathcal{M}= \frac{(m_1 m_2)^{3/5}}{(m_1+m_2)^{1/5}}\,,
\ee
with $m_{1,2}$ the masses of the two black holes in the binary. We focus on the case of an inspiraling BBH, for which the emitted power is given by \cite{Maggiore:1900zz} 
\begin{equation}
\frac{\dd E_s}{\dd t_s}  =\frac{32}{5 G c^5 } (\pi G \M f_s )^{10/3}=\frac{32}{5 G c^5 } (\pi G \M (1+z)f )^{10/3}\,,\label{eq:dEdlogf}
\end{equation}
where $E_s$ is the source-frame energy of one source with mass $\M$ and (source-frame) frequency $f_s$, and $t_s$ is the local cosmic time at the source. The source-frame frequency $f_s$ is related to the observed frequency by $f = f_s/(1+z)$. The observed power is found using $E_o = E_s/(1+z)$ and $\dd t_o = \dd t_s (1+z)$, hence
\be
\frac{\dd E_o}{\dd t_o} = \frac{1}{(1+z)^2}\frac{\dd E_s}{\dd t_s}\,.
\ee

We denote the (comoving) number of sources per comoving volume $\dd V_c$, per frequency $\dd f$ and per logarithmic chirp mass $\dd\log{\mathcal{M}}$ by
\be
a^3 n_s(\bn, z, \M, f ) \equiv f \frac{\dd N_s}{\dd V_c\, \dd \log(\M)\, \dd f} \propto f^{-8/3}\,,
\ee
where $N_s$ is the number of sources. The previous scaling in frequency is found using that the number of sources in a given frequency band is proportional to the time spent in that band, along with $\dd \log f/\dd t \propto f^{8/3}$. We split this quantity into an angle-averaged term $\bar{n}_s$ and an anisotropic contribution 
\be
n_s(\bn, z,\mathcal M, f)=\bar{n}_s(z,\mathcal M, f)+\delta n_s(\bn, z,\mathcal M, f)\, .
\ee

 To consider the contributions of the different directions $\bn$ to the total energy density parameter at the observer, we then analogously split $\Omega_{\rm GW}$ in an angle-averaged term $\bar\Omega_{\rm GW}$ and an anisotropic fluctuation, with the definition~\eqref{Splitbackgroundperturbations}. 
The isotropic part is obtained by summing all contributions as in~\eqref{Omegaf0} with 
\be
\partial_r \bar{\Omega}_{\text{GW}}(f,r) =\int \dd \log \M \frac{\partial \bar \Omega_{\rm GW}(f,r,\M)}{\partial r \partial \log \M}\,,
\ee
where 
\be\label{dOmegabardrdlogM}
\frac{\partial \bar \Omega_{\rm GW}(f,r,\M)}{\partial r \partial \log \M}  = \frac{1}{\rho_c} \frac{1}{(1+z)^2} \frac{\dd E_s}{\dd t_s} a^3 \bar{n}_s(z,\M,f)\,,
\ee
and the function $z=z(r)$ must be used to relate a given redshift $z$ to the conformal distance $r$.

Denoting the fractional fluctuation in the number density of sources by 
\be
\delta_s(\bn, z, \mathcal{M})=\frac{\delta n_s(\bn, z, \mathcal{M}, f)}{\bar{n}_s(z, \mathcal{M}, f)}\,,
\label{eq:deltas}
\ee
and using that the GW sources are biased tracers of the underlying distribution of galaxies, 
\be
\delta_s(\bn, z, \mathcal{M})=b_{s}(z, \mathcal{M})\delta_g(\bn, z)\, , \label{eq:def_bias}
\ee
we obtain 
\begin{align}
\delta\Omega_{\rm GW}(f, \bn) &=\frac{1}{4\pi}\int \dd r\int\dd \log\M \nn\\
&\times\frac{\partial \bar \Omega_{\rm GW}(f,r,\M)} {\partial r \partial \log \M} b_s(z,\M) \, \delta_g(\bn, z)\,,
\label{eq:deltaOmega}
\end{align}
where the integration on the conformal distance $r$ can be replaced by an integral on redshifts using the background cosmology and $\dd r = \dd z/H(z)$.

Eq.\,\eqref{eq:deltaOmega} relates the anisotropies in the energy density of an astrophysical SGWB to the anisotropies of the large-scale structure where SGWB sources are found. 
We assume that GW sources follow the same clustering structure as the galaxies, hence we take $b_s=1$ in our analysis.

\section{Energy density and shot noise from a catalogue}\label{Appcatalog}

For a given source $i$, we introduce the following quantity $P_i \equiv (\dd E_o/\dd t_o)_i/\rho_c = (\dd E_s/\dd t_s)_i/\rho_c/(1+z_i)^2$, such that for a collection of $N$ sources in directions $\bn_i$
\be
\Omega_{\rm GW}(\bn) = \sum_{i=1}^N \left(\frac{P_i}{4\pi r_i^2}\right) \delta^2(\bn,\bn_i)\,.
\ee
Using
\be
\delta^2(\bn_1,\bn_2)=\sum_{\ell m} Y_{\ell m}(\bn_1) Y^\star_{\ell m}(\bn_2)\,,
\ee
and an expansion $\Omega_{\rm GW}(\bn) = \sum_{\ell m}\Omega_{\ell m}^{\rm GW} Y_{\ell m}(\bn)$ we get
\be
\Omega^{\rm GW}_{\ell m} = \sum_{i=1}^N \left(\frac{P_i}{4\pi r_i^2}\right) Y^\star_{\ell m}(\bn_i)\,.
\ee
An estimator of the $C^{\rm GW}_\ell$ defined in~\eqref{ClGW} is
\be
\hat{C}^{\rm GW}_\ell = \frac{1}{2\ell+1}\sum_{m=-\ell}^\ell \Omega^{\rm GW}_{\ell m}\Omega^{{\rm GW}\star}_{\ell m}\,,
\ee
and for random positions of the sources, that is sources which are not located according to clustering properties, it is non-vanishing only due to the Poissonian nature of the source distribution. We want to average this estimator over the direction realizations of the sources, that is over random realizations of the set of directions $\{ \bn_i\}$, so as to estimate the shot noise. We shall use that (see e.g. \cite{Allen:2024mtn} or appendix D of \cite{Pitrou:2024scp})
\begin{equation}
\langle Y_{\ell m}(\bn_{i_1})\rangle_{\{\bn_i\}} =\frac{\delta_\ell^0 \delta_m^0}{\sqrt{4 \pi}}\,,
\end{equation}
and for $(\ell,m) \neq (0,0)$ 
\be
\langle Y_{\ell m}(\bn_{i_1})Y^\star_{\ell' m'}(\bn_{i_1})\rangle_{\{\bn_i\}} = \frac{\delta_{\ell \ell'} \delta_{m m'}}{4 \pi}\,.
\ee
Then when averaging over all random positions of these $N$ sources
\begin{align}
4\pi\langle \Omega_{\rm GW}(\bn)\rangle_{\{\bn_i\}} &= \sqrt{4 \pi} \langle \Omega^{\rm GW}_{00}\rangle_{\{\bn_i\}}=\sum_{i=1}^N \frac{P_i}{4\pi r_i^2}\,,\label{OmegaAverageDiscrete}\\
\langle \hat{C}^{\rm GW}_\ell \rangle_{\{ \bn_i \}} &= \frac{1}{4\pi}\sum_{i=1}^N \left(\frac{P_i}{4\pi r_i^2}\right)^2\,.\label{shotcatalog}
\end{align}
We identify the average GW density as $\bar{\Omega}_{\rm GW}=4\pi\langle \Omega_{\rm GW}(\bn)\rangle_{\{\bn_i\}}$ and shot noise with $N_{\rm shot}^{\rm GW}=\langle \hat{C}^{\rm GW}_\ell \rangle_{\{ \bn_i \}}$.

In order to check the consistency with the continuous integral formulas we use
\be
\sum_{i=1}^N \to \int \dd \log f \int \dd \log \M \int \dd r  (4\pi r^2 a^3 \bar{n}_s)\,,
\ee
and indeed applying this replacement rule to~\eqref{OmegaAverageDiscrete} we get~\eqref{dOmegabardrdlogM}.
Also for shot noise we get
\begin{align}
\langle \hat{C}^{\rm GW}_\ell \rangle_{\{ \bn_i \}} &\simeq \int \dd \log f \int \dd \log \M \int \dd r a^3 \bar{n}_s r^2  \left(\frac{P}{4\pi r^2}\right)^2\\
&\simeq \int \dd \log f \int \dd \log \M \int \dd r \frac{a^3 \bar{n}_s}{(4\pi)^2 r^2} \left(\frac{\dd E_s/\dd t_s}{(1+z)^2\rho_c}\right)^2\,,\nonumber
\end{align}
and using \eqref{dOmegabardrdlogM} we recover~\eqref{shots}. Hence when using  a catalog,  we have to use~\eqref{OmegaAverageDiscrete} to estimate the background GW density, and~\eqref{shotcatalog} to estimate the shot noise.

\bibliography{HDbib}

\end{document}